\documentclass[%
 reprint,
superscriptaddress,
 amsmath,amssymb,
 aps,prl
]{revtex4-1}

\usepackage{graphicx}
\usepackage{dcolumn}
\usepackage{bm}

\begin{document}

\preprint{APS/123-QED}

\title{Extended Interaction Length Laser-Driven Acceleration\\in a Tunable Dielectric Structure
}

\author{Sophie Crisp}
 \affiliation{Department of Physics and Astronomy, UCLA, Los Angeles, CA 90095, USA}
\author{Alexander Ody}%
 \affiliation{Department of Physics and Astronomy, UCLA, Los Angeles, CA 90095, USA}
 \author{Joel England}%
 \affiliation{SLAC National Accelerator Laboratory, Menlo Park, CA 94025, USA}
 \author{Pietro Musumeci}%
 \affiliation{Department of Physics and Astronomy, UCLA, Los Angeles, CA 90095, USA}

\date{\today}

\begin{abstract}
The development of long, tunable structures is critical to increasing energy gain in laser-driven dielectric accelerators (DLAs). Here we combine pulse-front-tilt illumination with slab-geometry structures assembled by precisely aligning off-the-shelf 4~mm long transmission gratings to achieve up to 200 keV energy modulation for 6 MeV injected electrons. The effective interaction length is longer than 1~mm, limited by dephasing of the accelerated particles in the structure. The piezo-based independent mounting system for the gratings allows tuning of the gap and field distribution inside the structure.
\end{abstract}

\maketitle

Shrinking accelerators to the optical scale holds the promise to reduce cost and increase availability of relativistic electron beams for scientific, industrial and medical applications \cite{England:RMP}. Leveraging the high damage threshold of dielectric materials as well as continuous progress in high power laser and nanofabrication technologies, laser-driven structure accelerators (or dielectric laser accelerators, DLAs) have already demonstrated GeV/m level gradients \cite{Cesar2018}, much larger than conventional accelerators and current research efforts are directed towards extending the interaction region. Notably, the technical challenges to achieve this goal are common to all advanced accelerators, including the plasma-based schemes \cite{esarey2009}, and are related to the physical dimensions of the accelerator (length of structure or plasma cell), the temporal walk-off associated with the different velocities of the drive pulse and the electron bunch (group velocity mismatch) and the loss of phase-synchronicity as the particles gain energy (dephasing). Depletion of the energy in the driving pulse then poses the fundamental limit to the acceleration length. 

Experimental demonstration of DLA acceleration has been accomplished using two main structure types: pillars and gratings. Dual pillar structures can be fabricated on single Si wafers, easing the nm-scale fabrication tolerances. They can then be illuminated (from the side or the top) without propagating high intensity laser pulses inside thick dielectric substrates \cite{Leedle2015a,Black2019,Schonenberger2019,Niedermayer2021,Shiloh2021}. Dual grating structures can have much larger aspect ratios,  a built-in collimation function which is useful to isolate the transmitted electrons, and can be made out of fused silica and/or coated with higher damage threshold materials \cite{Breuer2013,Leedle2015,Chlouba2022,Miao2020,Wootton2016,Cesar2018,Cesar2018PFT,Crisp2021}. Reaching long interaction lengths in both of these structures has been impeded by the constraints imposed by power delivery geometry. The highest accelerating gradients are only accessible using 100~fs laser pulses, which allow for high intensities while still remaining below the damage thresholds for most materials. Since the laser is typically coupled orthogonally to the direction of electron travel, the interaction length is set by its pulse length to the tens of $\mu$m scale \cite{Peralta2013}.  To overcome this limitation, a pulse-front-tilt (PFT) configuration has been employed, extending the interaction beyond the temporal laser envelope duration \cite{Wei2017,Kozak2018}, leading to the demonstration of 315~keV energy gain over a 700~$\mu$m interaction length \cite{Cesar2018PFT}. 

Additional energy gain requires manufacturing longer structures, stretching the state-of-the-art in nanofabrication techniques to meet tolerances for sustaining acceleration and preserving alignment over meso-scale (mm to cm) dimensions. Some degree of post-fabrication tuning would greatly ease these challenges and allow for more flexibility in the structure design. This is also due to the fact that efficiently interacting over a longer distance requires mitigating the loss of phase synchronicity (or resonant condition) caused by the particles gaining energy in the structure. Dephasing can be compensated, as recently shown in sub-relativistic experiments, by carefully chirping the parameters of the structure along its length. However, this limits the structure to a unique input beam energy and laser gradient. Resonant acceleration can also be preserved by the so called soft-tuning approach which entails control of electron dynamics through software based manipulation of the drive laser phase and is very appealing for its experimental flexibility \cite{shiloh2022}.

In this experiment we demonstrate the use of independently mounted commercial transmission gratings to form a 4~mm long dual grating structure for laser-driven acceleration. This structure is illuminated on a single side by a 2~mJ 780~nm, 100~fs laser in a PFT configuration and fed by the high brightness 6~MeV electron beam from the UCLA Pegasus photoinjector. By mounting the gratings on separate piezo controls, we can adjust the gap and the relative tooth offset to optimize the amplitude and symmetry of the fields experienced by the electrons and maximize in situ the energy modulation up to 200 keV. In agreement with FTDT simulations and optical characterization of the structure, a periodic slowly decaying relation between energy gain and gap size is observed. From the saturation of the energy gain for varying PFT laser sizes, a DLA interaction length of $>$ 1~mm is observed, short of the physical dimension of the grating, but fully consistent with dephasing in an unchirped structure. These results provide the first demonstration of an in situ tunable grating structure and also the longest DLA interaction to date. They represent a critical step forward in increasing the energy gain in DLA schemes to the MeV scale. 

\begin{figure}[h!]
    \centering
    \includegraphics[width =\columnwidth]{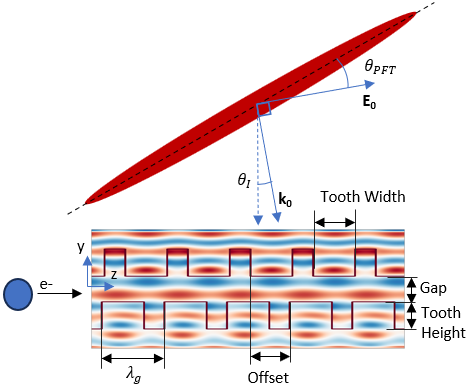}
    \caption{A cartoon showing a linearly polarized laser with pulse-front-tilt angle $\theta_{PFT}$ arriving at incident angle $\theta_I$ on a structure with periodicity $\lambda_g$. The field distribution of the evanescent mode inside the structure is shown.}
    \label{fig:grating_popout}
\end{figure}

The experimental geometry is shown in Fig. \ref{fig:grating_popout}, with a laser incident upon two parallel transmission gratings. The fields in the vacuum gap of an infinitely wide (no x-dependence) dual grating structure of period $\lambda_g = 2\pi/k_g$ illuminated by a laser of angular frequency $\omega$ and amplitude $E_0$ polarized along the direction of the electrons (in order to excite a TM wave) can be written as a sum of Floquet modes
\begin{equation}
\begin{array}{cc}
    E_{n,z} = E_0\left({d_n e^{-\Gamma_n y} + c_n e^{\Gamma_n y}}\right)e^{i(k_nz-\omega t)}\\
 \end{array}
 \label{eq:fields}
\end{equation}
where the $n_{th}$ mode, described by longitudinal wave number, $k_n = n k_g$ 
has normalized phase velocity $\beta_n = k_0/k_n$.

For phase-synchronous acceleration we focus our attention on the resonant mode (usually $n=1$) for which $\beta_n = \beta$. In order to satisfy Maxwell's equations, the transverse wavenumber is $\Gamma_n = \sqrt{k_n^2-k_0^2} = k_n/\gamma$. The complex parameters $d_n$ and $c_n$ depend on the mode number $n$, the input laser frequency, and the structure geometry and can be interpreted as the amplitudes of the waves diffracted from the top and bottom grating respectively.

\begin{figure}
    \centering
    \includegraphics[width = \columnwidth]{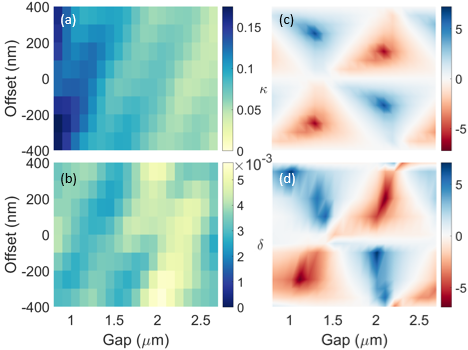}
    \caption{Dual grating structure parameters have strong dependence on offset and gap. Upper left) structure factor, $\kappa$, is seen to decay significantly with increased gap. Bottom left) Deflection factor, $\delta$; only where the deflection is near zero will electrons be transmitted. Note the amplitude of deflection is two orders of magnitude weaker than the acceleration force. Bottom right) Measured ratio of $\pm$ 1 diffraction order amplitudes from the assembled structure illuminated by a 635 nm diode laser (lower right) and corresponding FDTD simulations (upper right).}
    \label{fig:accelDeflecDiffrac}
\end{figure}

In a symmetrically illuminated structure, $c_n = d_n$ and $E_z$ could be described by a cosh-like mode centered in the middle of the structure gap. However, for this single-side illumination, it is instead described by the sum of a cosh-like and sinh-like mode. The structure factor, $\kappa$, is proportional to the acceleration gradient and can be written in terms of only the the parameters $c_n$ and $d_n$ as $\kappa_n = |d_n + c_n|$. In the upper left panel of Fig. \ref{fig:accelDeflecDiffrac} FTDT simulations results show how $\kappa$ decreases with increasing gap size as expected from the evanescent nature of the fields. A weaker, but clearly visible dependence on the relative offset between the teeth is also observed. 

Likewise, we can define a deflection parameter, $\delta$, which is the magnitude of the sinh mode within the structure. This deflection is proportional to $(k_{n}-k_0\beta)/\Gamma = 1/\gamma$, and is therefore defined as $\delta = |d_n - c_n|/\gamma$. The Fig. \ref{fig:accelDeflecDiffrac}b shows $\delta$ as a function of offset and gap for the gratings in this experiment. From this, it is clear that the deflection force can be minimized via correctly aligning the structure geometry, even with defined grating parameters. Note that deflection forces are 2 orders of magnitude smaller than the acceleration force, regardless of alignment. This allows the structures to have some angular misalignment while maintaining high throughput.

\begin{figure*}
    \centering
    \includegraphics[width = \textwidth]{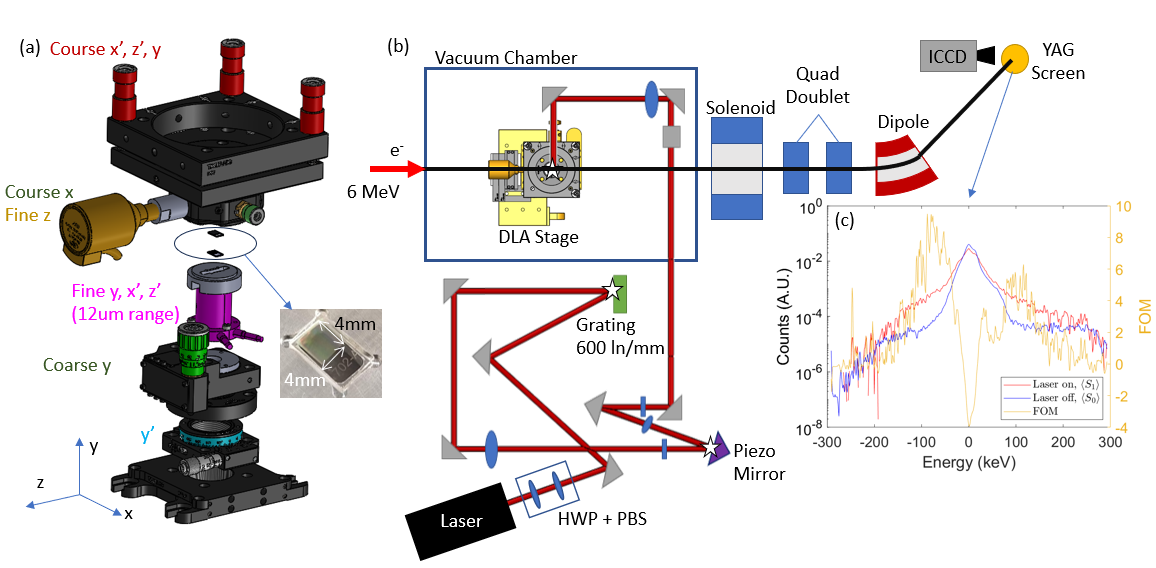}
    \caption{(a) The components of the mounting system; each colored component corresponds to a tunable component. Only the piezo motor, highlighted in pink, is controllable when the structure is in vacuum. The grating itself is shown in the small inset. (b) Experimental setup (not to scale). The UCLA Pegasus gun and linac generates 6~MeV electrons which are then focused into the DLA aperture by a quadrupole triplet and then sent to a dipole spectrometer where the beam is observed on a YAG screen, imaged by a gated intensified CCD camera (ICCD). The pulse-front-tilt optics are also shown, with imaging planes denoted by stars at the initial grating, an intermediate plane where the piezo motor controlled mirror is installed, and at the DLA. Cylindrical lenses are used to tune the intensity. (c) Typical laser on and laser off electron energy spectrum.}
    \label{fig:beamlineSchematic}
\end{figure*}

We use 4~mm square gratings etched on a 625 $\mu$m thick fused silica substrate with a tooth height of 855~nm, 65\% duty cycle, and 800~nm periodicity. The increased stiffness compared to thinner substrates is useful to avoid bending. Early attempts at bonding two wafers over multiple millimeters resulted in warped structures with micron-scale gap variability, so we opted for developing an independent mounting system. The gratings are mounted in a cage system with both coarse and fine controls of relative angle, gap, and offset, shown in Fig. \ref{fig:beamlineSchematic}a. The lower grating is glued at three points to its respective mount attached to a 3-axis vacuum-compatible piezo stage. 

Structures are characterized optically before beamline insertion using a 635~nm diode laser. During assembly of the structure, we first eliminate spatial thin film interference fringes using course angular adjustment followed by piezo fine tuning to flatten the gap. At this point, a tunable etalon effect on the reflectivity of the structure can be verified by changing the gap by $\lambda/2$. Once the gap is flat and small ($<6$~$\mu$m), interference in the  diffraction lobes can be used to set the relative grating rotation to near zero. Finally, the relative intensity of the first order diffraction lobes is recorded as a function of gap and offset. Simulations performed in Lumerical are compared to these measurements to retrieve the offset and gap  \cite{Crisp2022}, as shown in Fig. \ref{fig:accelDeflecDiffrac}.

The optical system makes use of a 20~mJ, 100~fs, 780~nm laser split 9 to 1 between a frequency tripling UV path for the photocathode and a drive line utilizing a pulse front tilt (PFT) configuration incident on the DLA. The PFT setup is similar to the one described by Cesar et al \cite{Cesar2018a} with an additional intermediate imaging plane where a piezo-controlled mirror can be used to adjust the angle of incidence on the DLA without changing the spatial alignment with the electron beam. A 600~ln/mm grating is followed by a 300~mm focal length achromatic lens to create the mid-point imaging plane. Two achromatic lenses (150~mm and 300~mm focal lengths respectively) are then used to precisely adjust the magnification and imaging plane location at the DLA structure. Since the DLA grating period (800~nm) is longer than the laser wavelength (780~nm), the phase matching condition $k_g - \frac{\omega_l}{c\beta} + \frac{\omega_l}{c}sin(\theta_I) = 0$ for a 6~MeV electron beam is satisfied by an incident angle $\theta_I$ = 28.1~mrad \cite{Crisp2021} which can be first set by careful alignment of the DLA backreflection and then tuned in with the piezo controlled mirror.

The overall magnification ($M = \frac{\tan(\theta_{PFT})}{d\lambda_l}  = 2.08$) is determined by group velocity matching the laser pulse to the electrons, $\beta = \frac{\cos(\theta_{PFT})}{\sin(\theta_I+\theta_{PFT})}$, yielding  $\theta_{PFT} = 44.3\deg$. The main laser PFT angle is directly measured to be 44.2$\pm 0.3 \deg$ over an interaction longer than $4$~mm by observing the location of the interference fringes as a function of relative time of arrival of a probe reference laser pulse at the DLA plane. Two additional cylindrical lenses are used to adjust the transverse laser spot size in the non-PFT dimension and control the fluence at the interaction.

A 1 pC, 6~MeV, 1~ps electron bunch is generated by the UCLA Pegasus gun and linac \cite{Alesini2015}, and focused at the DLA plane to an rms spot size of 100~$\mu$m with a normalized emittance of 200~nm. The measurement of the transmittance (approximately 1000 e-/shot with the laser off and a gap size of 1~$\mu$m) is consistent with these beam parameters and the structure dimensions. Note that in the initial setup we can take advantage of the piezo motor to widen the gap from 1~$\mu$m to 5~$\mu$m, and increase the transmission 26-fold, allowing for the optimization of pitch and yaw angle and e-beam spot size before decreasing the gap size. Downstream of the DLA, the beam is then transported to a dipole spectrometer, as shown in Fig. \ref{fig:beamlineSchematic}b. 

After overlapping spatially and temporally \cite{Scoby2013} the electron and laser beams at the DLA plane, first experiments are conducted using a reference flat laser pulse by replacing the nominal 600 ln/mm PFT grating with a mirror. In this case, the interaction length is set by the laser pulse length and the only quantities affecting energy modulation are the incident fluence and the structure factor. Once a modulation signal is stably obtained in the flat pulse case, we first replace the grating with a 1200 ln/mm (giving a PFT angle of approximately 62.8 degrees) to increase the interaction region to 240 $\mu$m and then go to the nominal grating and change the dimension of the laser along the pulse front tilt dimension to maximize interaction length.

In Fig. \ref{fig:beamlineSchematic}c we show a representative energy spectrum showing the highest energy modulation recorded in the experiment. The asymmetry in energy gain and loss is consistent with the angle of incidence in this particular case being lower than the resonant angle 28.1~mrad \cite{Cesar2018PFT}. In general, in order to analyze the spectra, we define the figure of merit as $FOM$ = $\frac{\langle S_1\rangle- \langle S_0\rangle}{\sigma_0}$ where $\langle S_{1 (0)} \rangle$ are the average of the observed spectra at least 5 laser on (off) shots and $\sigma_0$ is the standard deviation of the laser off shots. To better discriminate the signal at the tails of the spectrum where the electron density is low, we require $>$5 consecutive points on the data sets to have a signal-to-noise ratio larger than 1.25. We find the maximal energy gain from this experiment to be 200~keV.

\begin{figure}
\centering\includegraphics[width = \columnwidth]{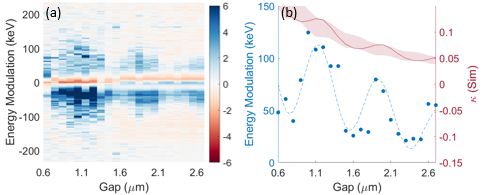}
\caption{Scan over gap size. (a) $FOM$ vs. gap between the transmission gratings. (b) Maximum energy modulation as a function of gap size extracted from (a). The simulated structure factor, $\kappa$, as function of gap is also shown. The filled red region represents the variation of $\kappa$ depending on the teeth offset. A sinusoidal fit with a decaying amplitude is fit to the data.}
\label{fig:GapScan}
\end{figure}

Our setup allows us for the first time to study the
performances of the DLA accelerator as a function of the gap between the gratings.  In Fig. \ref{fig:GapScan}, we show the results of the gap-scan at constant offset as performed by controlling the in-vacuum piezo motors. In agreement with simulation, we observe a clear decrease in DLA acceleration as the gap increases which can be explained by the lower structure factor. The shaded area in the figure shows the range of possible $\kappa$ depending on the teeth offset which is a parameter that can not be measured directly during the experiment. A particular line corresponding to an offset of 0~nm can be well matched to the data. In addition, while the depletion in the zero-loss main peak is evident and nearly constant at all gaps, the population of ac(de)celerated electrons changes in a periodic fashion, resulting in the energy modulation signal to vanish at certain gaps. This can be explained by considering the periodic variation of the deflection forces when adjusting the gap at constant offset (i.e. moving on an horizontal line in fig. \ref{fig:accelDeflecDiffrac}b.). Whenever the deflection forces are the strongest, no accelerated particles can make it through the narrow gap and the acceleration signal is lost. A sinusoidal fit with a decaying amplitude is overlaid to the data to take this effect into account.  

\begin{figure}
    \centering
    \includegraphics[width = \columnwidth]{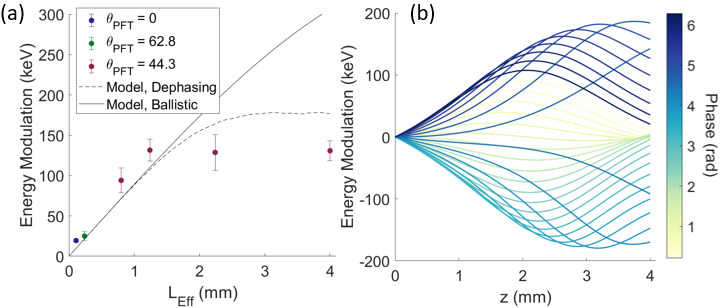}
    \caption{(a) Data shows increasing energy modulation up to an effective length of 1.24~mm. (b) Particle trajectories throughout a 4~mm DLA interaction. Particles are color-coded according to their initial phase. Due to dephasing, saturation of the energy gain occurs before the full length of the structure.}
    \label{fig:InteractionLength}
\end{figure}

We can also study the acceleration in this uniform dual grating structure for different interaction lengths. Fig. \ref{fig:InteractionLength}a shows the results where the interaction length is controlled in two distinct ways, i.e. by i) temporally or ii) spatially varying the overlap of the laser and the electrons at the DLA. The former is accomplished by swapping the PFT grating to create different $\theta_{PFT}$. In the $\theta_{PFT} = 62.8^{\circ}$ case, this amounts to a 2.2x longer interaction length than the flat pulse case, designated in the green point in fig. \ref{fig:InteractionLength}a.  Once the PFT angle is matched to the electron velocity, however, the interaction is instead limited by the spatial extent of the laser which can be controlled by placing a slit aperture just before the PFT grating imaging plane.  Fig. \ref{fig:InteractionLength}a shows the energy modulation increasing up to an interaction length of 1.24~mm and subsequently saturating. 

There are a number of reasons that could contribute to this plateau, including a slightly unmatched PFT angle, a spatial variation in the laser phase profile and a poor alignment of the electron beam and laser propagation axes. Nevertheless, even when accounting for these factors, the energy exchange would still be limited by the particle dephasing along the interaction, since the gratings are not tapered. In order to understand this effect we look at the energy of a particle in the DLA fields as calculated by simply integrating the field amplitude and taking into account the dynamical evoluation of the electron phase for given pulse front tilt and incident angles. We plot particle trajectories assuming an initial  6~MeV beam in Fig. \ref{fig:InteractionLength}b for different input phases. The trajectories demonstrate that electrons do not gain energy linearly over the full length of the grating structure. For example, electrons which enter the DLA at the optimal $2~\pi$ phase reach their peak energy around the center of the structure and are subsequently decelerated by the end of the DLA. We consider this to be the main reason leading to the plateau of the maximal energy gain within the structure. The impact of dephasing on the DLA longitudinal dynamics can be minimized by increasing the input electron energy as a stiffer the beam can resonantly interact for a longer distance in a structure of constant periodicity. In the non-relativistic regime, chirped structures which taper the structure period for continuous phase matching have been shown effective to mitigate this effect. 

In conclusion, we demonstrated the use of off-the-shelf gratings to accelerate electrons over a record 1.24~mm effective length. The use of commercial gratings to assemble a tunable DLA presents an attractive pathway to large-scale DLA development. The observed 200 keV energy modulation yields an average acceleration gradient of 0.16 GeV/m, mainly due to the non-optimized structure factor of the gratings. Further improvements can also be obtained by increasing the incident laser intensity, limited in the experiment by the low damage threshold of the grating antireflection coating layer. 

The piezo-controlled independent mounting system allowed for the first time for beam-based tuning of structure parameters and the accelerator performances. In particular, sub-micron accuracy in controlling the gap size over 4 mm length was demonstrated in both optical measurement and acceleration experiments, addressing the challenge of aligning nanostructure on the multi-mm mesoscale, which is a fundamental step towards increasing the energy gain for relativistic applications of the DLA acceleration scheme.  

In future experiments, the PFT optical setup used here will be modified to include a spatial light modulator to phase match the laser field to the accelerated electrons to mitigate the dephasing that caused saturation, allowing for multi-mm DLA acceleration lengths and energy gains of $>1$ MeV to be fully realized.
\begin{acknowledgments}
Thanks to K. Buchwald from Ibsen Photonics for his help finding commercial gratings. This work has been supported by the ACHIP grant from the Gordon and Betty Moore Foundation (GBMF4744) and by U.S. Department of Energy Grant No. DE-AC02-76SF00515. This material is based upon work supported by the National Science Foundation Graduate Research Fellowship Program under Grant No. DGE-1650604. Any opinions, findings, and conclusions or recommendations expressed in this material are those of the author(s) and do not necessarily reflect the views of the National Science Foundation.
\end{acknowledgments}


\begin{thebibliography}{23}%
\makeatletter
\providecommand \@ifxundefined [1]{%
 \@ifx{#1\undefined}
}%
\providecommand \@ifnum [1]{%
 \ifnum #1\expandafter \@firstoftwo
 \else \expandafter \@secondoftwo
 \fi
}%
\providecommand \@ifx [1]{%
 \ifx #1\expandafter \@firstoftwo
 \else \expandafter \@secondoftwo
 \fi
}%
\providecommand \natexlab [1]{#1}%
\providecommand \enquote  [1]{``#1''}%
\providecommand \bibnamefont  [1]{#1}%
\providecommand \bibfnamefont [1]{#1}%
\providecommand \citenamefont [1]{#1}%
\providecommand \href@noop [0]{\@secondoftwo}%
\providecommand \href [0]{\begingroup \@sanitize@url \@href}%
\providecommand \@href[1]{\@@startlink{#1}\@@href}%
\providecommand \@@href[1]{\endgroup#1\@@endlink}%
\providecommand \@sanitize@url [0]{\catcode `\\12\catcode `\$12\catcode
  `\&12\catcode `\#12\catcode `\^12\catcode `\_12\catcode `\%12\relax}%
\providecommand \@@startlink[1]{}%
\providecommand \@@endlink[0]{}%
\providecommand \url  [0]{\begingroup\@sanitize@url \@url }%
\providecommand \@url [1]{\endgroup\@href {#1}{\urlprefix }}%
\providecommand \urlprefix  [0]{URL }%
\providecommand \Eprint [0]{\href }%
\providecommand \doibase [0]{https://doi.org/}%
\providecommand \selectlanguage [0]{\@gobble}%
\providecommand \bibinfo  [0]{\@secondoftwo}%
\providecommand \bibfield  [0]{\@secondoftwo}%
\providecommand \translation [1]{[#1]}%
\providecommand \BibitemOpen [0]{}%
\providecommand \bibitemStop [0]{}%
\providecommand \bibitemNoStop [0]{.\EOS\space}%
\providecommand \EOS [0]{\spacefactor3000\relax}%
\providecommand \BibitemShut  [1]{\csname bibitem#1\endcsname}%
\let\auto@bib@innerbib\@empty
\bibitem [{\citenamefont {England}\ \emph {et~al.}(2014)\citenamefont
  {England}, \citenamefont {Noble}, \citenamefont {Bane}, \citenamefont
  {Dowell}, \citenamefont {Ng}, \citenamefont {Spencer}, \citenamefont
  {Tantawi}, \citenamefont {Wu}, \citenamefont {Byer}, \citenamefont {Peralta},
  \citenamefont {Soong}, \citenamefont {Chang}, \citenamefont {Montazeri},
  \citenamefont {Wolf}, \citenamefont {Cowan}, \citenamefont {Dawson},
  \citenamefont {Gai}, \citenamefont {Hommelhoff}, \citenamefont {Huang},
  \citenamefont {Jing}, \citenamefont {McGuinness}, \citenamefont {Palmer},
  \citenamefont {Naranjo}, \citenamefont {Rosenzweig}, \citenamefont {Travish},
  \citenamefont {Mizrahi}, \citenamefont {Schachter}, \citenamefont {Sears},
  \citenamefont {Werner},\ and\ \citenamefont {Yoder}}]{England:RMP}%
  \BibitemOpen
  \bibfield  {author} {\bibinfo {author} {\bibfnamefont {R.~J.}\ \bibnamefont
  {England}}, \bibinfo {author} {\bibfnamefont {R.~J.}\ \bibnamefont {Noble}},
  \bibinfo {author} {\bibfnamefont {K.}~\bibnamefont {Bane}}, \bibinfo {author}
  {\bibfnamefont {D.~H.}\ \bibnamefont {Dowell}}, \bibinfo {author}
  {\bibfnamefont {C.-K.}\ \bibnamefont {Ng}}, \bibinfo {author} {\bibfnamefont
  {J.~E.}\ \bibnamefont {Spencer}}, \bibinfo {author} {\bibfnamefont
  {S.}~\bibnamefont {Tantawi}}, \bibinfo {author} {\bibfnamefont
  {Z.}~\bibnamefont {Wu}}, \bibinfo {author} {\bibfnamefont {R.~L.}\
  \bibnamefont {Byer}}, \bibinfo {author} {\bibfnamefont {E.}~\bibnamefont
  {Peralta}}, \bibinfo {author} {\bibfnamefont {K.}~\bibnamefont {Soong}},
  \bibinfo {author} {\bibfnamefont {C.-M.}\ \bibnamefont {Chang}}, \bibinfo
  {author} {\bibfnamefont {B.}~\bibnamefont {Montazeri}}, \bibinfo {author}
  {\bibfnamefont {S.~J.}\ \bibnamefont {Wolf}}, \bibinfo {author}
  {\bibfnamefont {B.}~\bibnamefont {Cowan}}, \bibinfo {author} {\bibfnamefont
  {J.}~\bibnamefont {Dawson}}, \bibinfo {author} {\bibfnamefont
  {W.}~\bibnamefont {Gai}}, \bibinfo {author} {\bibfnamefont {P.}~\bibnamefont
  {Hommelhoff}}, \bibinfo {author} {\bibfnamefont {Y.-C.}\ \bibnamefont
  {Huang}}, \bibinfo {author} {\bibfnamefont {C.}~\bibnamefont {Jing}},
  \bibinfo {author} {\bibfnamefont {C.}~\bibnamefont {McGuinness}}, \bibinfo
  {author} {\bibfnamefont {R.~B.}\ \bibnamefont {Palmer}}, \bibinfo {author}
  {\bibfnamefont {B.}~\bibnamefont {Naranjo}}, \bibinfo {author} {\bibfnamefont
  {J.}~\bibnamefont {Rosenzweig}}, \bibinfo {author} {\bibfnamefont
  {G.}~\bibnamefont {Travish}}, \bibinfo {author} {\bibfnamefont
  {A.}~\bibnamefont {Mizrahi}}, \bibinfo {author} {\bibfnamefont
  {L.}~\bibnamefont {Schachter}}, \bibinfo {author} {\bibfnamefont
  {C.}~\bibnamefont {Sears}}, \bibinfo {author} {\bibfnamefont {G.~R.}\
  \bibnamefont {Werner}},\ and\ \bibinfo {author} {\bibfnamefont {R.~B.}\
  \bibnamefont {Yoder}},\ }\bibfield  {title} {\bibinfo {title} {Dielectric
  laser accelerators},\ }\href {https://doi.org/10.1103/RevModPhys.86.1337}
  {\bibfield  {journal} {\bibinfo  {journal} {Rev. Mod. Phys.}\ }\textbf
  {\bibinfo {volume} {86}},\ \bibinfo {pages} {1337} (\bibinfo {year}
  {2014})}\BibitemShut {NoStop}%
\bibitem [{\citenamefont {Cesar}\ \emph
  {et~al.}(2018{\natexlab{a}})\citenamefont {Cesar}, \citenamefont {Custodio},
  \citenamefont {Maxson}, \citenamefont {Musumeci}, \citenamefont {Shen},
  \citenamefont {Threlkeld}, \citenamefont {England}, \citenamefont {Hanuka},
  \citenamefont {Makasyuk}, \citenamefont {Peralta}, \citenamefont {Wootton},\
  and\ \citenamefont {Wu}}]{Cesar2018}%
  \BibitemOpen
  \bibfield  {author} {\bibinfo {author} {\bibfnamefont {D.}~\bibnamefont
  {Cesar}}, \bibinfo {author} {\bibfnamefont {S.}~\bibnamefont {Custodio}},
  \bibinfo {author} {\bibfnamefont {J.}~\bibnamefont {Maxson}}, \bibinfo
  {author} {\bibfnamefont {P.}~\bibnamefont {Musumeci}}, \bibinfo {author}
  {\bibfnamefont {X.}~\bibnamefont {Shen}}, \bibinfo {author} {\bibfnamefont
  {E.}~\bibnamefont {Threlkeld}}, \bibinfo {author} {\bibfnamefont {R.~J.}\
  \bibnamefont {England}}, \bibinfo {author} {\bibfnamefont {A.}~\bibnamefont
  {Hanuka}}, \bibinfo {author} {\bibfnamefont {I.~V.}\ \bibnamefont
  {Makasyuk}}, \bibinfo {author} {\bibfnamefont {E.~A.}\ \bibnamefont
  {Peralta}}, \bibinfo {author} {\bibfnamefont {K.~P.}\ \bibnamefont
  {Wootton}},\ and\ \bibinfo {author} {\bibfnamefont {Z.}~\bibnamefont {Wu}},\
  }\bibfield  {title} {\bibinfo {title} {{High-field nonlinear optical response
  and phase control in a dielectric laser accelerator}},\ }\href
  {https://doi.org/10.1038/s42005-018-0047-y} {\bibfield  {journal} {\bibinfo
  {journal} {Communications Physics 2018 1:1}\ }\textbf {\bibinfo {volume}
  {1}},\ \bibinfo {pages} {1} (\bibinfo {year}
  {2018}{\natexlab{a}})}\BibitemShut {NoStop}%
\bibitem [{\citenamefont {Esarey}\ \emph {et~al.}(2009)\citenamefont {Esarey},
  \citenamefont {Schroeder},\ and\ \citenamefont {Leemans}}]{esarey2009}%
  \BibitemOpen
  \bibfield  {author} {\bibinfo {author} {\bibfnamefont {E.}~\bibnamefont
  {Esarey}}, \bibinfo {author} {\bibfnamefont {C.~B.}\ \bibnamefont
  {Schroeder}},\ and\ \bibinfo {author} {\bibfnamefont {W.~P.}\ \bibnamefont
  {Leemans}},\ }\bibfield  {title} {\bibinfo {title} {Physics of laser-driven
  plasma-based electron accelerators},\ }\href@noop {} {\bibfield  {journal}
  {\bibinfo  {journal} {Reviews of modern physics}\ }\textbf {\bibinfo {volume}
  {81}},\ \bibinfo {pages} {1229} (\bibinfo {year} {2009})}\BibitemShut
  {NoStop}%
\bibitem [{\citenamefont {Leedle}\ \emph
  {et~al.}(2015{\natexlab{a}})\citenamefont {Leedle}, \citenamefont {Ceballos},
  \citenamefont {Deng}, \citenamefont {Solgaard}, \citenamefont {Pease},
  \citenamefont {Byer},\ and\ \citenamefont {Harris}}]{Leedle2015a}%
  \BibitemOpen
  \bibfield  {author} {\bibinfo {author} {\bibfnamefont {K.~J.}\ \bibnamefont
  {Leedle}}, \bibinfo {author} {\bibfnamefont {A.}~\bibnamefont {Ceballos}},
  \bibinfo {author} {\bibfnamefont {H.}~\bibnamefont {Deng}}, \bibinfo {author}
  {\bibfnamefont {O.}~\bibnamefont {Solgaard}}, \bibinfo {author}
  {\bibfnamefont {R.~F.}\ \bibnamefont {Pease}}, \bibinfo {author}
  {\bibfnamefont {R.~L.}\ \bibnamefont {Byer}},\ and\ \bibinfo {author}
  {\bibfnamefont {J.~S.}\ \bibnamefont {Harris}},\ }\bibfield  {title}
  {\bibinfo {title} {{Dielectric laser acceleration of sub-100 keV electrons
  with silicon dual-pillar grating structures}},\ }\href
  {https://doi.org/10.1364/OL.40.004344} {\bibfield  {journal} {\bibinfo
  {journal} {Optics Letters, Vol. 40, Issue 18, pp. 4344-4347}\ }\textbf
  {\bibinfo {volume} {40}},\ \bibinfo {pages} {4344} (\bibinfo {year}
  {2015}{\natexlab{a}})}\BibitemShut {NoStop}%
\bibitem [{\citenamefont {Black}\ \emph {et~al.}(2019)\citenamefont {Black},
  \citenamefont {Niedermayer}, \citenamefont {Miao}, \citenamefont {Zhao},
  \citenamefont {Solgaard}, \citenamefont {Byer},\ and\ \citenamefont
  {Leedle}}]{Black2019}%
  \BibitemOpen
  \bibfield  {author} {\bibinfo {author} {\bibfnamefont {D.~S.}\ \bibnamefont
  {Black}}, \bibinfo {author} {\bibfnamefont {U.}~\bibnamefont {Niedermayer}},
  \bibinfo {author} {\bibfnamefont {Y.}~\bibnamefont {Miao}}, \bibinfo {author}
  {\bibfnamefont {Z.}~\bibnamefont {Zhao}}, \bibinfo {author} {\bibfnamefont
  {O.}~\bibnamefont {Solgaard}}, \bibinfo {author} {\bibfnamefont {R.~L.}\
  \bibnamefont {Byer}},\ and\ \bibinfo {author} {\bibfnamefont {K.~J.}\
  \bibnamefont {Leedle}},\ }\bibfield  {title} {\bibinfo {title} {{Net
  Acceleration and Direct Measurement of Attosecond Electron Pulses in a
  Silicon Dielectric Laser Accelerator}},\ }\href
  {https://doi.org/10.1103/PHYSREVLETT.123.264802/FIGURES/3/MEDIUM} {\bibfield
  {journal} {\bibinfo  {journal} {Physical Review Letters}\ }\textbf {\bibinfo
  {volume} {123}},\ \bibinfo {pages} {264802} (\bibinfo {year}
  {2019})}\BibitemShut {NoStop}%
\bibitem [{\citenamefont {Sch\"onenberger}\ \emph {et~al.}(2019)\citenamefont
  {Sch\"onenberger}, \citenamefont {Mittelbach}, \citenamefont {Yousefi},
  \citenamefont {McNeur}, \citenamefont {Niedermayer},\ and\ \citenamefont
  {Hommelhoff}}]{Schonenberger2019}%
  \BibitemOpen
  \bibfield  {author} {\bibinfo {author} {\bibfnamefont {N.}~\bibnamefont
  {Sch\"onenberger}}, \bibinfo {author} {\bibfnamefont {A.}~\bibnamefont
  {Mittelbach}}, \bibinfo {author} {\bibfnamefont {P.}~\bibnamefont {Yousefi}},
  \bibinfo {author} {\bibfnamefont {J.}~\bibnamefont {McNeur}}, \bibinfo
  {author} {\bibfnamefont {U.}~\bibnamefont {Niedermayer}},\ and\ \bibinfo
  {author} {\bibfnamefont {P.}~\bibnamefont {Hommelhoff}},\ }\bibfield  {title}
  {\bibinfo {title} {Generation and characterization of attosecond microbunched
  electron pulse trains via dielectric laser acceleration},\ }\href
  {https://doi.org/10.1103/PhysRevLett.123.264803} {\bibfield  {journal}
  {\bibinfo  {journal} {Phys. Rev. Lett.}\ }\textbf {\bibinfo {volume} {123}},\
  \bibinfo {pages} {264803} (\bibinfo {year} {2019})}\BibitemShut {NoStop}%
\bibitem [{\citenamefont {Niedermayer}\ \emph {et~al.}(2021)\citenamefont
  {Niedermayer}, \citenamefont {Black}, \citenamefont {Leedle}, \citenamefont
  {Miao}, \citenamefont {Byer},\ and\ \citenamefont
  {Solgaard}}]{Niedermayer2021}%
  \BibitemOpen
  \bibfield  {author} {\bibinfo {author} {\bibfnamefont {U.}~\bibnamefont
  {Niedermayer}}, \bibinfo {author} {\bibfnamefont {D.~S.}\ \bibnamefont
  {Black}}, \bibinfo {author} {\bibfnamefont {K.~J.}\ \bibnamefont {Leedle}},
  \bibinfo {author} {\bibfnamefont {Y.}~\bibnamefont {Miao}}, \bibinfo {author}
  {\bibfnamefont {R.~L.}\ \bibnamefont {Byer}},\ and\ \bibinfo {author}
  {\bibfnamefont {O.}~\bibnamefont {Solgaard}},\ }\bibfield  {title} {\bibinfo
  {title} {Low-energy-spread attosecond bunching and coherent electron
  acceleration in dielectric nanostructures},\ }\href
  {https://doi.org/10.1103/PhysRevApplied.15.L021002} {\bibfield  {journal}
  {\bibinfo  {journal} {Phys. Rev. Appl.}\ }\textbf {\bibinfo {volume} {15}},\
  \bibinfo {pages} {L021002} (\bibinfo {year} {2021})}\BibitemShut {NoStop}%
\bibitem [{\citenamefont {Shiloh}\ \emph {et~al.}(2021)\citenamefont {Shiloh},
  \citenamefont {Chlouba}, \citenamefont {Yousefi},\ and\ \citenamefont
  {Hommelhoff}}]{Shiloh2021}%
  \BibitemOpen
  \bibfield  {author} {\bibinfo {author} {\bibfnamefont {R.}~\bibnamefont
  {Shiloh}}, \bibinfo {author} {\bibfnamefont {T.}~\bibnamefont {Chlouba}},
  \bibinfo {author} {\bibfnamefont {P.}~\bibnamefont {Yousefi}},\ and\ \bibinfo
  {author} {\bibfnamefont {P.}~\bibnamefont {Hommelhoff}},\ }\bibfield  {title}
  {\bibinfo {title} {{Particle acceleration using top-illuminated nanophotonic
  dielectric structures}},\ }\href {https://doi.org/10.1364/OE.420235}
  {\bibfield  {journal} {\bibinfo  {journal} {Optics Express, Vol. 29, Issue
  10, pp. 14403-14411}\ }\textbf {\bibinfo {volume} {29}},\ \bibinfo {pages}
  {14403} (\bibinfo {year} {2021})}\BibitemShut {NoStop}%
\bibitem [{\citenamefont {Breuer}\ and\ \citenamefont
  {Hommelhoff}(2013)}]{Breuer2013}%
  \BibitemOpen
  \bibfield  {author} {\bibinfo {author} {\bibfnamefont {J.}~\bibnamefont
  {Breuer}}\ and\ \bibinfo {author} {\bibfnamefont {P.}~\bibnamefont
  {Hommelhoff}},\ }\bibfield  {title} {\bibinfo {title} {{Laser-based
  acceleration of nonrelativistic electrons at a dielectric structure}},\
  }\href {https://doi.org/10.1103/PhysRevLett.111.134803} {\bibfield  {journal}
  {\bibinfo  {journal} {Physical Review Letters}\ }\textbf {\bibinfo {volume}
  {111}},\ \bibinfo {pages} {134803} (\bibinfo {year} {2013})},\ \Eprint
  {https://arxiv.org/abs/1308.0464} {arXiv:1308.0464} \BibitemShut {NoStop}%
\bibitem [{\citenamefont {Leedle}\ \emph
  {et~al.}(2015{\natexlab{b}})\citenamefont {Leedle}, \citenamefont {Pease},
  \citenamefont {Byer},\ and\ \citenamefont {Harris}}]{Leedle2015}%
  \BibitemOpen
  \bibfield  {author} {\bibinfo {author} {\bibfnamefont {K.~J.}\ \bibnamefont
  {Leedle}}, \bibinfo {author} {\bibfnamefont {R.~F.}\ \bibnamefont {Pease}},
  \bibinfo {author} {\bibfnamefont {R.~L.}\ \bibnamefont {Byer}},\ and\
  \bibinfo {author} {\bibfnamefont {J.~S.}\ \bibnamefont {Harris}},\ }\bibfield
   {title} {\bibinfo {title} {{Laser acceleration and deflection of 96.3 keV
  electrons with a silicon dielectric structure}},\ }\href
  {https://doi.org/10.1364/OPTICA.2.000158} {\bibfield  {journal} {\bibinfo
  {journal} {Optica, Vol. 2, Issue 2, pp. 158-161}\ }\textbf {\bibinfo {volume}
  {2}},\ \bibinfo {pages} {158} (\bibinfo {year}
  {2015}{\natexlab{b}})}\BibitemShut {NoStop}%
\bibitem [{\citenamefont {Chlouba}\ \emph {et~al.}(2022)\citenamefont
  {Chlouba}, \citenamefont {Shiloh}, \citenamefont {Karlsson}, \citenamefont
  {Koz{\'{a}}k}, \citenamefont {Hommelhoff}, \citenamefont {Hamberg},\ and\
  \citenamefont {Forsberg}}]{Chlouba2022}%
  \BibitemOpen
  \bibfield  {author} {\bibinfo {author} {\bibfnamefont {T.}~\bibnamefont
  {Chlouba}}, \bibinfo {author} {\bibfnamefont {R.}~\bibnamefont {Shiloh}},
  \bibinfo {author} {\bibfnamefont {M.}~\bibnamefont {Karlsson}}, \bibinfo
  {author} {\bibfnamefont {M.}~\bibnamefont {Koz{\'{a}}k}}, \bibinfo {author}
  {\bibfnamefont {P.}~\bibnamefont {Hommelhoff}}, \bibinfo {author}
  {\bibfnamefont {M.}~\bibnamefont {Hamberg}},\ and\ \bibinfo {author}
  {\bibfnamefont {P.}~\bibnamefont {Forsberg}},\ }\bibfield  {title} {\bibinfo
  {title} {{Diamond-based dielectric laser acceleration}},\ }\href
  {https://doi.org/10.1364/OE.442752} {\bibfield  {journal} {\bibinfo
  {journal} {Optics Express, Vol. 30, Issue 1, pp. 505-510}\ }\textbf {\bibinfo
  {volume} {30}},\ \bibinfo {pages} {505} (\bibinfo {year} {2022})}\BibitemShut
  {NoStop}%
\bibitem [{\citenamefont {Miao}\ \emph {et~al.}(2020)\citenamefont {Miao},
  \citenamefont {Black}, \citenamefont {Leedle}, \citenamefont {Zhao},
  \citenamefont {Deng}, \citenamefont {Ceballos}, \citenamefont {Byer},
  \citenamefont {Harris},\ and\ \citenamefont {Solgaard}}]{Miao2020}%
  \BibitemOpen
  \bibfield  {author} {\bibinfo {author} {\bibfnamefont {Y.}~\bibnamefont
  {Miao}}, \bibinfo {author} {\bibfnamefont {D.~S.}\ \bibnamefont {Black}},
  \bibinfo {author} {\bibfnamefont {K.~J.}\ \bibnamefont {Leedle}}, \bibinfo
  {author} {\bibfnamefont {Z.}~\bibnamefont {Zhao}}, \bibinfo {author}
  {\bibfnamefont {H.}~\bibnamefont {Deng}}, \bibinfo {author} {\bibfnamefont
  {A.}~\bibnamefont {Ceballos}}, \bibinfo {author} {\bibfnamefont {R.~L.}\
  \bibnamefont {Byer}}, \bibinfo {author} {\bibfnamefont {J.~S.}\ \bibnamefont
  {Harris}},\ and\ \bibinfo {author} {\bibfnamefont {O.}~\bibnamefont
  {Solgaard}},\ }\bibfield  {title} {\bibinfo {title} {Surface treatments of
  dielectric laser accelerators for increased laser-induced damage threshold},\
  }\href {https://doi.org/10.1364/OL.379628} {\bibfield  {journal} {\bibinfo
  {journal} {Opt. Lett.}\ }\textbf {\bibinfo {volume} {45}},\ \bibinfo {pages}
  {391} (\bibinfo {year} {2020})}\BibitemShut {NoStop}%
\bibitem [{\citenamefont {Wootton}\ \emph {et~al.}(2016)\citenamefont
  {Wootton}, \citenamefont {Wu}, \citenamefont {Cowan}, \citenamefont {Hanuka},
  \citenamefont {Makasyuk}, \citenamefont {Peralta}, \citenamefont {Soong},
  \citenamefont {Byer},\ and\ \citenamefont {England}}]{Wootton2016}%
  \BibitemOpen
  \bibfield  {author} {\bibinfo {author} {\bibfnamefont {K.~P.}\ \bibnamefont
  {Wootton}}, \bibinfo {author} {\bibfnamefont {Z.}~\bibnamefont {Wu}},
  \bibinfo {author} {\bibfnamefont {B.~M.}\ \bibnamefont {Cowan}}, \bibinfo
  {author} {\bibfnamefont {A.}~\bibnamefont {Hanuka}}, \bibinfo {author}
  {\bibfnamefont {I.~V.}\ \bibnamefont {Makasyuk}}, \bibinfo {author}
  {\bibfnamefont {E.~A.}\ \bibnamefont {Peralta}}, \bibinfo {author}
  {\bibfnamefont {K.}~\bibnamefont {Soong}}, \bibinfo {author} {\bibfnamefont
  {R.~L.}\ \bibnamefont {Byer}},\ and\ \bibinfo {author} {\bibfnamefont
  {R.~J.}\ \bibnamefont {England}},\ }\bibfield  {title} {\bibinfo {title}
  {Demonstration of acceleration of relativistic electrons at a dielectric
  microstructure using femtosecond laser pulses},\ }\href
  {https://doi.org/10.1364/OL.41.002696} {\bibfield  {journal} {\bibinfo
  {journal} {Opt. Lett.}\ }\textbf {\bibinfo {volume} {41}},\ \bibinfo {pages}
  {2696} (\bibinfo {year} {2016})}\BibitemShut {NoStop}%
\bibitem [{\citenamefont {Cesar}\ \emph
  {et~al.}(2018{\natexlab{b}})\citenamefont {Cesar}, \citenamefont {Maxson},
  \citenamefont {Shen}, \citenamefont {Wootton}, \citenamefont {Tan},
  \citenamefont {England},\ and\ \citenamefont {Musumeci}}]{Cesar2018PFT}%
  \BibitemOpen
  \bibfield  {author} {\bibinfo {author} {\bibfnamefont {D.}~\bibnamefont
  {Cesar}}, \bibinfo {author} {\bibfnamefont {J.}~\bibnamefont {Maxson}},
  \bibinfo {author} {\bibfnamefont {X.}~\bibnamefont {Shen}}, \bibinfo {author}
  {\bibfnamefont {K.~P.}\ \bibnamefont {Wootton}}, \bibinfo {author}
  {\bibfnamefont {S.}~\bibnamefont {Tan}}, \bibinfo {author} {\bibfnamefont
  {R.~J.}\ \bibnamefont {England}},\ and\ \bibinfo {author} {\bibfnamefont
  {P.}~\bibnamefont {Musumeci}},\ }\bibfield  {title} {\bibinfo {title}
  {Enhanced energy gain in a dielectric laser accelerator using a tilted pulse
  front laser},\ }\href {https://doi.org/10.1364/OE.26.029216} {\bibfield
  {journal} {\bibinfo  {journal} {Opt. Express}\ }\textbf {\bibinfo {volume}
  {26}},\ \bibinfo {pages} {29216} (\bibinfo {year}
  {2018}{\natexlab{b}})}\BibitemShut {NoStop}%
\bibitem [{\citenamefont {Crisp}\ \emph {et~al.}(2021)\citenamefont {Crisp},
  \citenamefont {Ody}, \citenamefont {Musumeci},\ and\ \citenamefont
  {England}}]{Crisp2021}%
  \BibitemOpen
  \bibfield  {author} {\bibinfo {author} {\bibfnamefont {S.}~\bibnamefont
  {Crisp}}, \bibinfo {author} {\bibfnamefont {A.}~\bibnamefont {Ody}}, \bibinfo
  {author} {\bibfnamefont {P.}~\bibnamefont {Musumeci}},\ and\ \bibinfo
  {author} {\bibfnamefont {R.~J.}\ \bibnamefont {England}},\ }\bibfield
  {title} {\bibinfo {title} {Resonant phase matching by oblique illumination of
  a dielectric laser accelerator},\ }\href
  {https://doi.org/10.1103/PhysRevAccelBeams.24.121305} {\bibfield  {journal}
  {\bibinfo  {journal} {Phys. Rev. Accel. Beams}\ }\textbf {\bibinfo {volume}
  {24}},\ \bibinfo {pages} {121305} (\bibinfo {year} {2021})}\BibitemShut
  {NoStop}%
\bibitem [{\citenamefont {Peralta}\ \emph {et~al.}(2013)\citenamefont
  {Peralta}, \citenamefont {Soong}, \citenamefont {England}, \citenamefont
  {Colby}, \citenamefont {Wu}, \citenamefont {Montazeri}, \citenamefont
  {McGuinness}, \citenamefont {McNeur}, \citenamefont {Leedle}, \citenamefont
  {Walz}, \citenamefont {Sozer}, \citenamefont {Cowan}, \citenamefont
  {Schwartz}, \citenamefont {Travish},\ and\ \citenamefont
  {Byer}}]{Peralta2013}%
  \BibitemOpen
  \bibfield  {author} {\bibinfo {author} {\bibfnamefont {E.~A.}\ \bibnamefont
  {Peralta}}, \bibinfo {author} {\bibfnamefont {K.}~\bibnamefont {Soong}},
  \bibinfo {author} {\bibfnamefont {R.~J.}\ \bibnamefont {England}}, \bibinfo
  {author} {\bibfnamefont {E.~R.}\ \bibnamefont {Colby}}, \bibinfo {author}
  {\bibfnamefont {Z.}~\bibnamefont {Wu}}, \bibinfo {author} {\bibfnamefont
  {B.}~\bibnamefont {Montazeri}}, \bibinfo {author} {\bibfnamefont
  {C.}~\bibnamefont {McGuinness}}, \bibinfo {author} {\bibfnamefont
  {J.}~\bibnamefont {McNeur}}, \bibinfo {author} {\bibfnamefont {K.~J.}\
  \bibnamefont {Leedle}}, \bibinfo {author} {\bibfnamefont {D.}~\bibnamefont
  {Walz}}, \bibinfo {author} {\bibfnamefont {E.~B.}\ \bibnamefont {Sozer}},
  \bibinfo {author} {\bibfnamefont {B.}~\bibnamefont {Cowan}}, \bibinfo
  {author} {\bibfnamefont {B.}~\bibnamefont {Schwartz}}, \bibinfo {author}
  {\bibfnamefont {G.}~\bibnamefont {Travish}},\ and\ \bibinfo {author}
  {\bibfnamefont {R.~L.}\ \bibnamefont {Byer}},\ }\bibfield  {title} {\bibinfo
  {title} {{Demonstration of electron acceleration in a laser-driven dielectric
  microstructure}},\ }\href {https://doi.org/10.1038/nature12664} {\bibfield
  {journal} {\bibinfo  {journal} {Nature 2013 503:7474}\ }\textbf {\bibinfo
  {volume} {503}},\ \bibinfo {pages} {91} (\bibinfo {year} {2013})}\BibitemShut
  {NoStop}%
\bibitem [{\citenamefont {Wei}\ \emph {et~al.}(2017)\citenamefont {Wei},
  \citenamefont {Ibison}, \citenamefont {Xia}, \citenamefont {Smith},\ and\
  \citenamefont {Welsch}}]{Wei2017}%
  \BibitemOpen
  \bibfield  {author} {\bibinfo {author} {\bibfnamefont {Y.}~\bibnamefont
  {Wei}}, \bibinfo {author} {\bibfnamefont {M.}~\bibnamefont {Ibison}},
  \bibinfo {author} {\bibfnamefont {G.}~\bibnamefont {Xia}}, \bibinfo {author}
  {\bibfnamefont {J.~D.~A.}\ \bibnamefont {Smith}},\ and\ \bibinfo {author}
  {\bibfnamefont {C.~P.}\ \bibnamefont {Welsch}},\ }\bibfield  {title}
  {\bibinfo {title} {Dual-grating dielectric accelerators driven by a
  pulse-front-tilted laser},\ }\href {https://doi.org/10.1364/AO.56.008201}
  {\bibfield  {journal} {\bibinfo  {journal} {Appl. Opt.}\ }\textbf {\bibinfo
  {volume} {56}},\ \bibinfo {pages} {8201} (\bibinfo {year}
  {2017})}\BibitemShut {NoStop}%
\bibitem [{\citenamefont {Kozák}\ \emph {et~al.}(2018)\citenamefont {Kozák},
  \citenamefont {McNeur}, \citenamefont {Schönenberger}, \citenamefont
  {Illmer}, \citenamefont {Li}, \citenamefont {Tafel}, \citenamefont {Yousefi},
  \citenamefont {Eckstein},\ and\ \citenamefont {Hommelhoff}}]{Kozak2018}%
  \BibitemOpen
  \bibfield  {author} {\bibinfo {author} {\bibfnamefont {M.}~\bibnamefont
  {Kozák}}, \bibinfo {author} {\bibfnamefont {J.}~\bibnamefont {McNeur}},
  \bibinfo {author} {\bibfnamefont {N.}~\bibnamefont {Schönenberger}},
  \bibinfo {author} {\bibfnamefont {J.}~\bibnamefont {Illmer}}, \bibinfo
  {author} {\bibfnamefont {A.}~\bibnamefont {Li}}, \bibinfo {author}
  {\bibfnamefont {A.}~\bibnamefont {Tafel}}, \bibinfo {author} {\bibfnamefont
  {P.}~\bibnamefont {Yousefi}}, \bibinfo {author} {\bibfnamefont
  {T.}~\bibnamefont {Eckstein}},\ and\ \bibinfo {author} {\bibfnamefont
  {P.}~\bibnamefont {Hommelhoff}},\ }\bibfield  {title} {\bibinfo {title}
  {{Ultrafast scanning electron microscope applied for studying the interaction
  between free electrons and optical near-fields of periodic nanostructures}},\
  }\href {https://doi.org/10.1063/1.5032093} {\bibfield  {journal} {\bibinfo
  {journal} {Journal of Applied Physics}\ }\textbf {\bibinfo {volume} {124}},\
  \bibinfo {pages} {023104} (\bibinfo {year} {2018})},\ \Eprint
  {https://arxiv.org/abs/https://pubs.aip.org/aip/jap/article-pdf/doi/10.1063/1.5032093/14864034/023104\_1\_online.pdf}
  {https://pubs.aip.org/aip/jap/article-pdf/doi/10.1063/1.5032093/14864034/023104\_1\_online.pdf}
  \BibitemShut {NoStop}%
\bibitem [{\citenamefont {Shiloh}\ \emph {et~al.}(2022)\citenamefont {Shiloh},
  \citenamefont {Sch{\"o}nenberger}, \citenamefont {Adiv}, \citenamefont
  {Ruimy}, \citenamefont {Karnieli}, \citenamefont {Hughes}, \citenamefont
  {England}, \citenamefont {Leedle}, \citenamefont {Black}, \citenamefont
  {Zhao} \emph {et~al.}}]{shiloh2022}%
  \BibitemOpen
  \bibfield  {author} {\bibinfo {author} {\bibfnamefont {R.}~\bibnamefont
  {Shiloh}}, \bibinfo {author} {\bibfnamefont {N.}~\bibnamefont
  {Sch{\"o}nenberger}}, \bibinfo {author} {\bibfnamefont {Y.}~\bibnamefont
  {Adiv}}, \bibinfo {author} {\bibfnamefont {R.}~\bibnamefont {Ruimy}},
  \bibinfo {author} {\bibfnamefont {A.}~\bibnamefont {Karnieli}}, \bibinfo
  {author} {\bibfnamefont {T.}~\bibnamefont {Hughes}}, \bibinfo {author}
  {\bibfnamefont {R.~J.}\ \bibnamefont {England}}, \bibinfo {author}
  {\bibfnamefont {K.~J.}\ \bibnamefont {Leedle}}, \bibinfo {author}
  {\bibfnamefont {D.~S.}\ \bibnamefont {Black}}, \bibinfo {author}
  {\bibfnamefont {Z.}~\bibnamefont {Zhao}}, \emph {et~al.},\ }\bibfield
  {title} {\bibinfo {title} {Miniature light-driven nanophotonic electron
  acceleration and control},\ }\href@noop {} {\bibfield  {journal} {\bibinfo
  {journal} {Advances in Optics and Photonics}\ }\textbf {\bibinfo {volume}
  {14}},\ \bibinfo {pages} {862} (\bibinfo {year} {2022})}\BibitemShut
  {NoStop}%
\bibitem [{\citenamefont {Crisp}\ \emph {et~al.}(2022)\citenamefont {Crisp},
  \citenamefont {Musumeci},\ and\ \citenamefont {Ody}}]{Crisp2022}%
  \BibitemOpen
  \bibfield  {author} {\bibinfo {author} {\bibfnamefont {S.}~\bibnamefont
  {Crisp}}, \bibinfo {author} {\bibfnamefont {P.}~\bibnamefont {Musumeci}},\
  and\ \bibinfo {author} {\bibfnamefont {A.}~\bibnamefont {Ody}},\ }\bibfield
  {title} {\bibinfo {title} {All optical chartacterization of a dual grating
  accelerator structure},\ }in\ \href
  {https://doi.org/10.18429/JACoW-IPAC2022-WEOXSP2} {\emph {\bibinfo
  {booktitle} {Proc. IPAC'22}}},\ \bibinfo {series and number} {\bibinfo
  {series} {International Particle Accelerator Conference}\ No.~\bibinfo
  {number} {13}}\ (\bibinfo  {publisher} {JACoW Publishing, Geneva,
  Switzerland},\ \bibinfo {year} {2022})\ pp.\ \bibinfo {pages}
  {1602--1605}\BibitemShut {NoStop}%
\bibitem [{\citenamefont {Cesar}\ \emph
  {et~al.}(2018{\natexlab{c}})\citenamefont {Cesar}, \citenamefont {Maxson},
  \citenamefont {Musumeci}, \citenamefont {Shen}, \citenamefont {England},\
  and\ \citenamefont {Wootton}}]{Cesar2018a}%
  \BibitemOpen
  \bibfield  {author} {\bibinfo {author} {\bibfnamefont {D.}~\bibnamefont
  {Cesar}}, \bibinfo {author} {\bibfnamefont {J.}~\bibnamefont {Maxson}},
  \bibinfo {author} {\bibfnamefont {P.}~\bibnamefont {Musumeci}}, \bibinfo
  {author} {\bibfnamefont {X.}~\bibnamefont {Shen}}, \bibinfo {author}
  {\bibfnamefont {R.~J.}\ \bibnamefont {England}},\ and\ \bibinfo {author}
  {\bibfnamefont {K.~P.}\ \bibnamefont {Wootton}},\ }\bibfield  {title}
  {\bibinfo {title} {{Optical design for increased interaction length in a high
  gradient dielectric laser accelerator}},\ }\href
  {https://doi.org/10.1016/J.NIMA.2018.01.012} {\bibfield  {journal} {\bibinfo
  {journal} {Nuclear Instruments and Methods in Physics Research Section A:
  Accelerators, Spectrometers, Detectors and Associated Equipment}\ }\textbf
  {\bibinfo {volume} {909}},\ \bibinfo {pages} {252} (\bibinfo {year}
  {2018}{\natexlab{c}})},\ \Eprint {https://arxiv.org/abs/1801.01115}
  {arXiv:1801.01115} \BibitemShut {NoStop}%
\bibitem [{\citenamefont {Alesini}\ \emph {et~al.}(2015)\citenamefont
  {Alesini}, \citenamefont {Battisti}, \citenamefont {Ferrario}, \citenamefont
  {Foggetta}, \citenamefont {Lollo}, \citenamefont {Ficcadenti}, \citenamefont
  {Pettinacci}, \citenamefont {Custodio}, \citenamefont {Pirez}, \citenamefont
  {Musumeci},\ and\ \citenamefont {Palumbo}}]{Alesini2015}%
  \BibitemOpen
  \bibfield  {author} {\bibinfo {author} {\bibfnamefont {D.}~\bibnamefont
  {Alesini}}, \bibinfo {author} {\bibfnamefont {A.}~\bibnamefont {Battisti}},
  \bibinfo {author} {\bibfnamefont {M.}~\bibnamefont {Ferrario}}, \bibinfo
  {author} {\bibfnamefont {L.}~\bibnamefont {Foggetta}}, \bibinfo {author}
  {\bibfnamefont {V.}~\bibnamefont {Lollo}}, \bibinfo {author} {\bibfnamefont
  {L.}~\bibnamefont {Ficcadenti}}, \bibinfo {author} {\bibfnamefont
  {V.}~\bibnamefont {Pettinacci}}, \bibinfo {author} {\bibfnamefont
  {S.}~\bibnamefont {Custodio}}, \bibinfo {author} {\bibfnamefont
  {E.}~\bibnamefont {Pirez}}, \bibinfo {author} {\bibfnamefont
  {P.}~\bibnamefont {Musumeci}},\ and\ \bibinfo {author} {\bibfnamefont
  {L.}~\bibnamefont {Palumbo}},\ }\bibfield  {title} {\bibinfo {title} {New
  technology based on clamping for high gradient radio frequency photogun},\
  }\href {https://doi.org/10.1103/PhysRevSTAB.18.092001} {\bibfield  {journal}
  {\bibinfo  {journal} {Phys. Rev. ST Accel. Beams}\ }\textbf {\bibinfo
  {volume} {18}},\ \bibinfo {pages} {092001} (\bibinfo {year}
  {2015})}\BibitemShut {NoStop}%
\bibitem [{\citenamefont {Scoby}\ \emph {et~al.}(2013)\citenamefont {Scoby},
  \citenamefont {Li},\ and\ \citenamefont {Musumeci}}]{Scoby2013}%
  \BibitemOpen
  \bibfield  {author} {\bibinfo {author} {\bibfnamefont {C.~M.}\ \bibnamefont
  {Scoby}}, \bibinfo {author} {\bibfnamefont {R.~K.}\ \bibnamefont {Li}},\ and\
  \bibinfo {author} {\bibfnamefont {P.}~\bibnamefont {Musumeci}},\ }\bibfield
  {title} {\bibinfo {title} {{Effect of an ultrafast laser induced plasma on a
  relativistic electron beam to determine temporal overlap in pump–probe
  experiments}},\ }\href {https://doi.org/10.1016/J.ULTRAMIC.2012.07.015}
  {\bibfield  {journal} {\bibinfo  {journal} {Ultramicroscopy}\ }\textbf
  {\bibinfo {volume} {127}},\ \bibinfo {pages} {14} (\bibinfo {year}
  {2013})}\BibitemShut {NoStop}%
\end{thebibliography}

%

\end{document}